\documentclass[aps, prl, twocolumn, notitlepage,superscriptaddress,nofootinbib]{revtex4-1}

\usepackage{natbib}
\usepackage{graphicx}
\usepackage{dcolumn} 
\usepackage{bm}
\usepackage{hyperref}
\usepackage{mathtools}
\usepackage{amssymb}
\usepackage{enumitem}
\usepackage[mathlines]{lineno}
\usepackage[titletoc, title]{appendix}

\begin{document}

\title{Competing Correlated Insulators in multi-orbital systems coupled to phonons}

\author{Alberto Scazzola}
\affiliation{Scuola Internazionale Superiore di Studi Avanzati (SISSA), Via Bonomea 265, I-34136, Trieste, Italy}
\author{Adriano Amaricci}
\affiliation{CNR-IOM Democritos, Via Bonomea 265, I-34136 Trieste, Italy}
\author{Massimo Capone}
\affiliation{Scuola Internazionale Superiore di Studi Avanzati (SISSA), Via Bonomea 265, I-34136, Trieste, Italy}
\affiliation{CNR-IOM Democritos, Via Bonomea 265, I-34136 Trieste, Italy}

\date{\today}

\begin{abstract}
We study the interplay between electron-electron interaction and a Jahn-Teller phonon coupling in a two-orbital Hubbard model. We  demonstrate that the e-ph interaction coexists with the  Mott localization driven by the Hubbard repulsion $U$, but it competes with the Hund's coupling $J$. This interplay leads to two spectacularly different Mott insulators, a standard high-spin Mott insulator with frozen phonons which is stable when the Hund's coupling prevails, and a low-spin Mott-bipolaronic insulator favoured by phonons, where the characteristic features of Mott insulators and bipolarons coexist. 
The two phases are separated by a sharp boundary along which an intriguing intermediate solution emerges as a kind of compromise between the two solutions. 
\end{abstract}

\maketitle

Electron-phonon (e-ph) coupling and electron-electron (e-e) repulsion
are the key interactions determining the properties of electrons in
solids. Their relative importance can vary significantly across different
classes of materials, often leading to neglect one or the other in
the theoretical treatment.
On the other hand, it is nowadays clear that the interplay between e-e and e-ph interaction 
plays an important role in a wide class of materials, from colossal
magnetoresistance manganites~\cite{MillisDoubleExchangeAlone} to different
families of high-temperature
superconductors~\cite{Grilli1994,PhysRevB.69.144520,Kulic2000,Gunnarsson_2008,Capone2010,PhysRevLett.112.027002,PhysRevLett.119.107003,Capone_Science,Han2000,Nomura2015,Nomura2016}.

A wide majority of studies of the interplay between such two
interactions is based on a single-band Hubbard-Holstein (HH) model
description, featuring an on-site Hubbard repulsion and a coupling between the local electron density and phonons~\cite{Freericks1995,Capone1997,Koller2004,Capone2004,Sangiovanni2005,Paci2005,Perroni2005,Sangiovanni2006,Barone2006,Paci2006,Sangiovanni2006b,Macridin2006,Tezuka2007,Werner2007,Barone2008,DiCiolo2009,Bauer2010,Nowadnick2012,Murakami2013,Johnston2013,Hohenadler2013,Nowadnick2015,Imada2017,Karakuzu2017,Weber2018,Ghosh2018,Costa2020,Demler2020,Han2020}.
The simplicity of the HH model may lead to identify it as the paradigm to study the two interactions, but this expectation is somewhat misleading, i.e. that the properties of this model are not as generic as expected.  Indeed, in the HH model the phonons give rise to a 
frequency-dependent local attraction which has exactly the same form of the Hubbard repulsion. Thus the two interactions directly compete and the physics is essentially controlled by the comparison of the coupling strengths. The only non-trivial interplay arises 
because the $U$ term is instantaneous and the phonon term is retarded\cite{Sangiovanni2005}. 
A richer outcome can be realized in single-band systems considering
non-local e-ph couplings which compete less directly with the repulsion~\cite{Perroni2004,Casula2012,Giovannetti2014,Oelsen2010,Seibold2011,Sous2017}.

On the other hand, a single-band model is not sufficient to capture
the properties of many interesting materials, e.g. Fe-based
superconductors, Fullerides, Manganites, etc.
Multi-orbital models with e-e interaction have been widely
investigated recently, emphasizing the  non-trivial role of the Hund's 
coupling in the physics of strong correlations and its relevance for different materials~\cite{Georges_Hund,Janus}.
As far as the e-ph interaction is concerned, in multi-orbital systems the phonons can be coupled to local degrees of freedom different from the total charge, for example the occupation difference between orbitals or they hybridization. The interactions mediated by these phonons do not compete directly with the effect of the Hubbard repulsion, which freezes charge fluctuations~\cite{Capone_Science,Han2003,Capone2004,Capone2009}.

In this work we approach this physics solving a two-orbital model including a multi-orbital
e-e interaction, parameterized by the Hubbard $U$ and the Hund's
coupling $J$, and a Jahn-Teller (JT) e-ph mode. 
Using Dynamical Mean-Field Theory (DMFT)~\cite{DMFT} we focus on the
different strongly correlated phases originating from the interplay of
these two interactions. This study is not designed to describe any specific material, but rather to identify general trends to be used as a building block to study e-ph interaction in correlated materials, similarly to studies of the role of the Hund's coupling in models\cite{Georges_Hund} which have been instrumental to understand their role in iron-based superconductors and rhutenates.
Our results show that the JT phonon mode {\it cooperates} with the Hubbard interaction, but it competes with the Hund's exchange $J$. Indeed the key parameter to control the behavior of the system turns out to be the ratio between the e-ph and the
Hund's coupling.
The prevalence of either the Hund coupling or the JT phonon favours
different strongly correlated Mott-like phases with
remarkably distinct properties, i.e. a high-spin Mott
insulator and a low-spin Mott-bipolaronic state where phonon
fingerprints and Mott physics coexist.
Finally, we identify a hybrid Mott insulator resulting from a perfect
balance between the effects of the e-ph and the Hund's couplings.


We consider a two-orbital Hubbard model with two degenerate bands
described by a semicircular density of states with half-bandwidth
$D$. The e-e interaction is taken as
\begin{align}
\label{eq:H_ee}
H_{\mathrm{e-e}} & = U\sum_{i,a} n_{ia\uparrow}n_{ia\downarrow} 
+ (U-3J)\sum_{i,a<b, \, \sigma} n_{ia\sigma}n_{ib\sigma} \nonumber\\
& + (U-2J)\sum_{i,a\neq b} n_{ia\uparrow}n_{ib\downarrow}  
\end{align}
where $i$, $a$ and $b$, $\sigma$ are respectively site, orbital and
spin indices. This popular form of the interaction can
be seen as a Kanamori model without spin-flip and pair-hopping
terms.  
A dispersionless mode of frequency $\omega_0$ is coupled with the
difference in the occupation between the two orbitals, i.e. the
orbital polarization: 
\begin{align}
\label{eq:H_eph}
H_{\mathrm{e-ph}} & = g\sum_{i\sigma} (n_{i1\sigma} -n_{i2\sigma})(a_i^{\dagger} + a_i) + \omega_0 \sum_{i} a_i^{\dagger}a_i
\end{align}
This term corresponds to one of the two JT modes of a two-orbital
$e_g$ manifold (usually referred to as $Q_3$ and corresponding to an
orthorombic distortion which makes the z-axis inequivalent to the xy
plane). We did not include the other JT mode in order to keep the number of
parameters relatively small, while the Holstein coupling is expected to
reproduce the single-band results~\cite{Li2017,Li2018}. 


We set the density to half-filling, namely $n=2$, for which
the interactions can have qualitative effects, e.g., the e-e
interaction can lead to a Mott insulator. We define $\lambda = 2g^2/\omega_0$ as the strength of the
phonon-induced attraction.
 In order to identify the intrinsic correlation effects
we focus on paramagnetic phases without spatial, spin or orbital
ordering.

We solve the model using DMFT, which maps the lattice problem onto a
quantum impurity model subject to a self-consistency condition which
contains the information on the lattice model.
In our case, the impurity model features also a local phonon
mode. DMFT allows for a non-perturbative solution treating the two
interaction terms on the same footing without assuming any hierarchy
between the different energy scales of the problem.
We solve the impurity model using
exact-diagonalization~\cite{amaricci2021edipack,Caffarel1994,Capone2007}
after a truncation of the impurity model to a small number of levels,
that we typically take as $N_b = 8$, and to a limited number of
phononic levels, that we typically fix to $N^{ph}_{max} = 25$.
The convergence with respect to these parameters has been tested.

We characterize the metal-insulator transitions using the
quasiparticle weight $Z$ which measures the degree of metallicity of
the system,
the intra-orbital double occupation $\langle n_{im\uparrow}
n_{im\downarrow} \rangle$ with $m = 1,2$ and the local magnetic moment
measured by
$\langle S^2_z \rangle =\langle (\sum_{m}S_{z\,im})^2\rangle$.
All quantities do not depend on the site index $i$, thus in the
following we will drop reference to it.
We also consider specific phononic quantities, such as the number of
phonons $N_{ph} = \langle a^{\dagger} a\rangle$ and the phonon
distribution function $P(X)$ which measures the quantum amplitude that
the distortion operator of the phonon field assumes a value $X$. 

 \begin{figure}
   \centering
    \includegraphics[width=1\linewidth]{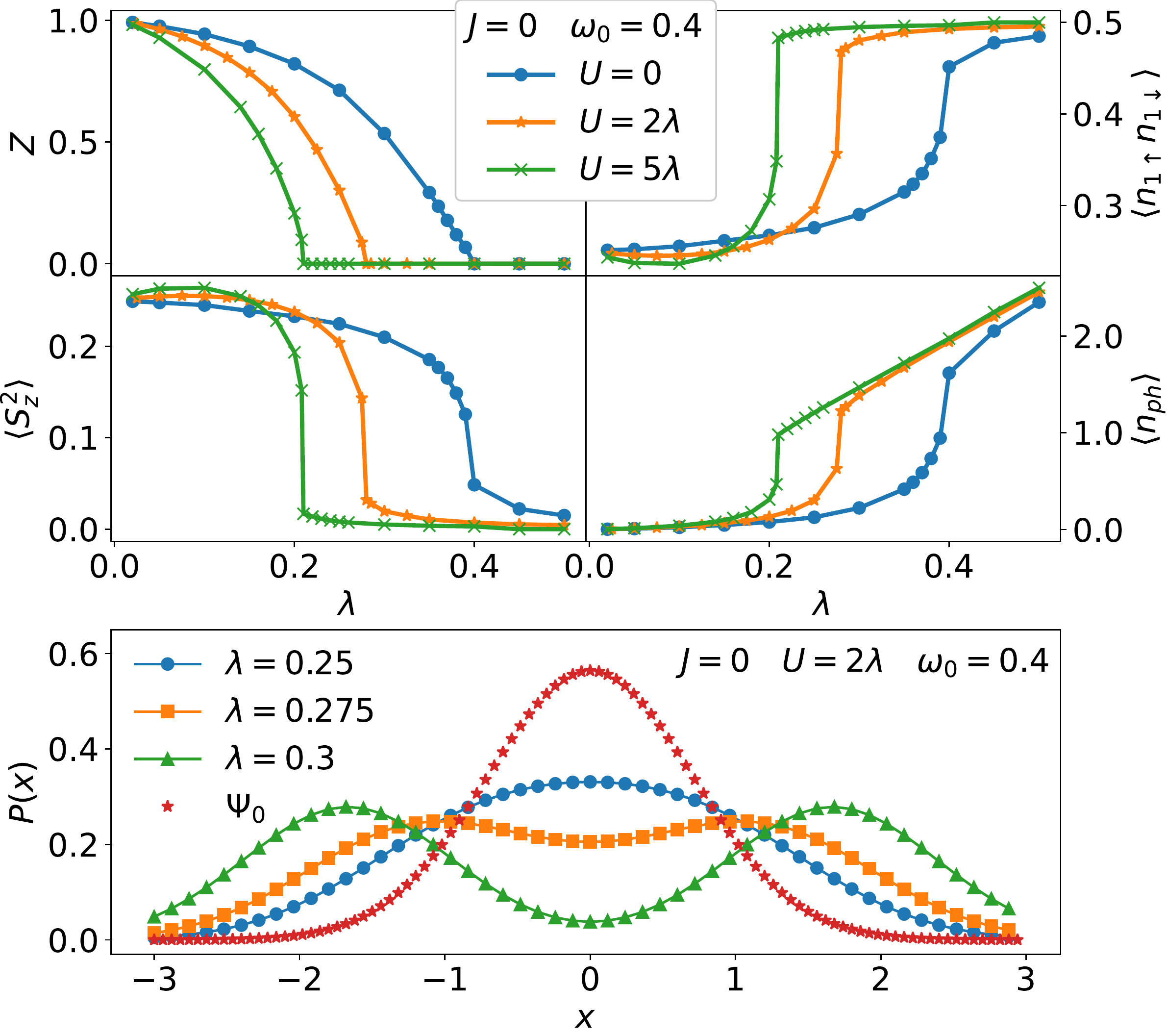}
    \caption{Approach to the bipolaronic-Mott insulator for $J=0$. In
      the top four panels we report the evolution of $Z$, $\langle
      n_{in\uparrow} n_{in\downarrow} \rangle$, $\langle S^2_z
      \rangle$ and $\langle n_{ph}\rangle$ as a function of $\lambda$ for three different values
      of $U/\lambda$ and $\omega_0/D = 0.4$.
      The bottom panel shows the evolution of the phonon distribution
      function $P(X)$ across the metal-insulator transition for
      $U/\lambda=2$ compared with that of an uncoupled phonon
      $\Psi_0$.}
    
    \label{fig:1}
\end{figure}

Our model depends on  four independent parameters $U/D$, $J/D$, $\lambda/D$ and $\omega_0/D$.
In the following we will keep the ratios $U/J$ and $\lambda/J$
constant when considering $J>0$,  while we vary the e-e interaction
strenght $U/D$.
This choice mimics the effect of pressure and chemical
substitution in a solid, which strongly change the hopping while leaving the
local interactions less affected. We take the ratio $\omega_0/D$ as an independent variable in order to
measure the effect of the degree of adiabaticity (retardation) of the
e-ph coupling.

We start our discussion from the limit $J=0$. In 
Fig.\ref{fig:1} we report various quantities as a function of $\lambda$ and three values of $U/\lambda$. For $U =0$ we have pure e-ph interaction.  Increasing $\lambda$ leads to a reduction of the quasiparticle weight $Z$ (enhancement of the effective mass) which eventually vanishes for $\lambda \simeq 0.4$ signaling a metal-insulator transition. The intra-orbital double occupancy evolves towards 0.5 while the squared magnetic moment falls towards zero. 
We conclude that we reached an insulating state where every site has
two electrons on one of the two orbitals, while the other is
empty. Half of the sites have two electrons on orbital 1, the other
half on orbital 2.
The number of phonons in the groundstate increases sharply as the
insulating state is approached. 
Importantly, the phonon distribution function $P(X)$ is centered
around $X=0$ in the whole metallic region, but it turns into a bimodal
with maxima at $\pm X_0 \neq 0$ at the transition, testifying that the
insulator is the two-orbital realization of a bipolaronic
state~\cite{Capone2003p,Capone2006p} where every lattice site has a
finite distortion, half with a sign and half with the other. The sites
where orbital 1(2) is doubly occupied have a negative(positive)
distortion. The picture does not depend qualitatively on $\omega_0/D$,
as we show in the supplementary information.

We now switch on the $U$ repulsion, while keeping $J=0$.  Since the
ratio $U/\lambda$ is kept constant, $U$ increases when we increase
$\lambda$. Remarkably, even if we consider a sizeable $U$ larger than
$\lambda$ ($U/\lambda =2$ and $5$ in Fig. \ref{fig:1}), the picture
obtained for $\lambda=0$ is not destroyed. Indeed all the observables
follow the  same qualitative behavior of $U=0$.
Even more surprisingly the Hubbard repulsion {\it {favours}} the
bipolaronic transition, which takes place for smaller $\lambda$ as
$U/\lambda$ increases. This is in sharp contrast with the naive
expectation based on the competition of  e-e and e-ph interactions and
the results for the Hubbard-Holstein model, for which a large $U$ completely quenches phononic effects~\cite{Sangiovanni2005}.

This unexpected result can be easily understood noticing that for
$J=0$ the interaction is $U/2\sum_i n_i^2$ (with
$n_i=\sum_{a\sigma}n_{ia\sigma}$) which simply selects
configurations with doubly occupied sites, i.e. configurations where
the local occupation equals the average, without any preference for
the internal arrangement between the orbitals.
If we denote by $\vert n_1,n_2\rangle$ configurations with $n_a$
electrons in the orbital $a=1,2$ we have that  
$\vert 0,2\rangle$ and $\vert 2,0\rangle$, which are favoured by the
phonons and characterize the bipolaronic state, are also compatible
with the effect of $U$.
This in turn helps the e-ph coupling to filter out configurations with
different local occupation. Thus, the Mott localization and the
bipolaron formation works in synergy to stabilize a  {\it{low-spin
    Mott-bipolaronic}} insulator. Such a state can not be realized in
a single-band model where Mott and bipolaronic insulators are mutually
exclusive.
\begin{figure}
    \centering
    \includegraphics[width=1\linewidth]{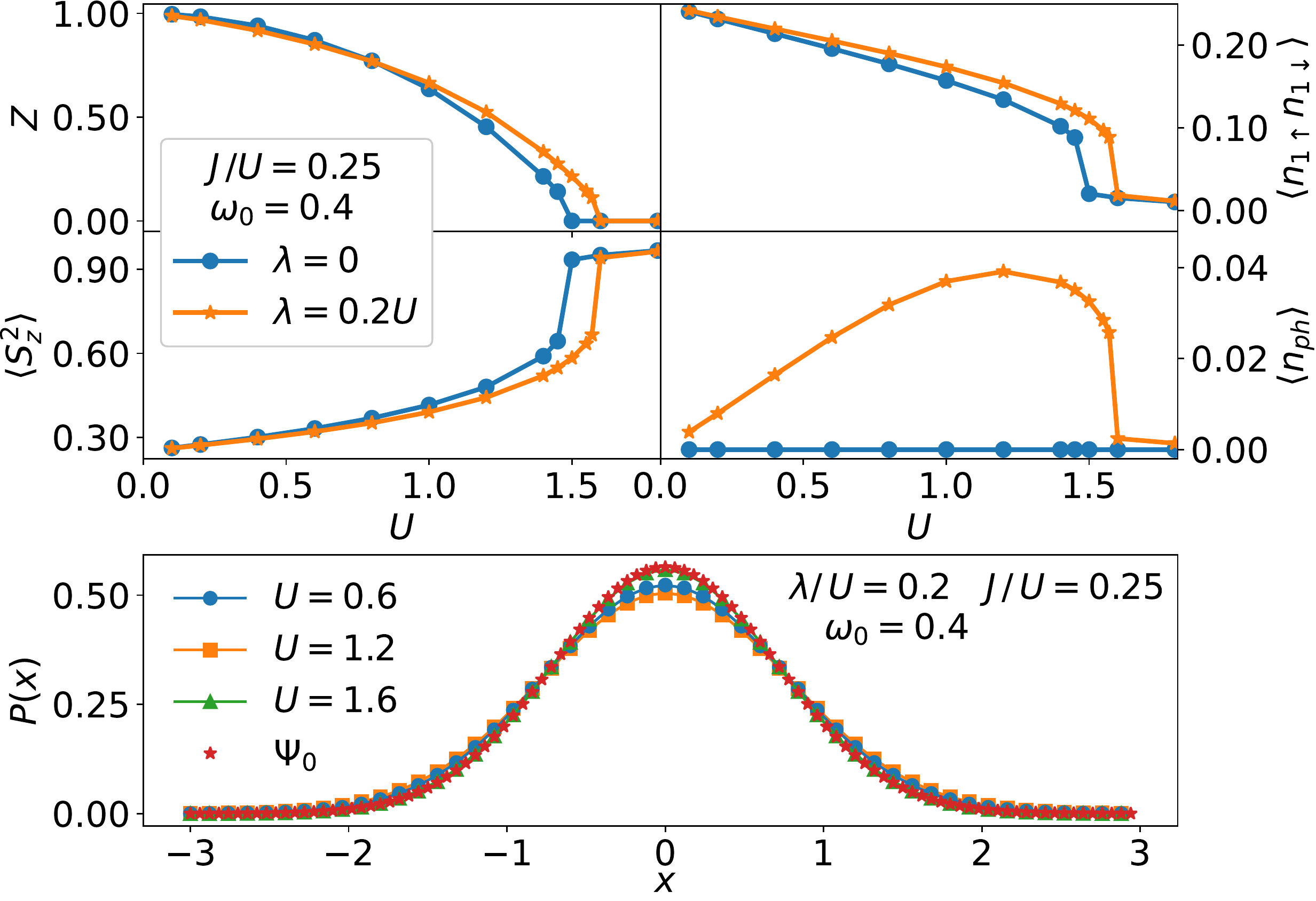}
    \caption{Approach to the high-spin Mott insulator for zero or
      small $\lambda$. In the top four panels we report the evolution
      of $Z$, $ \langle n_{m\uparrow} n_{m\downarrow} \rangle$,
      $\langle S^2_{z}\rangle$ and $\langle n_{ph}\rangle$ as a function of $U/D$ for two
      distinct values of $\lambda/U$ and $\omega_0/D = 0.4$.
      The bottom panel shows the evolution of the phonon distribution
      function $P(X)$ across the metal-insulator transition for
      $\lambda/U=0.2$ compared with that of an uncoupled phonon
      $\Psi_0$.
  }    
    \label{fig:2}
\end{figure}

The inclusion of the  Hund's coupling changes the picture by  favoring high-spin configurations
and disfavoring double occupation of the same orbital.
In Fig. \ref{fig:2} we show the same quantities as in \ref{fig:1} for
fixed ratio $J/U=0.25$, comparing the results for $\lambda =0$ with
those for $\lambda/U =0.2$.
Without e-ph coupling we recover a two-component Hubbard model.
In this case, increasing $U$ we induce a Mott transition signaled by a
vanishing $Z$. The presence of $J$ makes the Mott insulator
high-spin.
For density-density interaction, the Mott insulator mainly features
configurations with $S=1$ and $S_z = \pm 1$, completely different from
those characterizing the bipolaronic insulator discussed
above.
Accordingly, when $Z$ vanishes, the intra-orbital double occupation
tends to zero, while the magnetic moment increases towards 1.

The inclusion of a  moderate e-ph coupling ($\lambda/U =0.2$) 
does not alter the picture besides a small increase of the critical interaction
strength for the Mott transition.
The number of excited phonons is very small and eventually
drops to zero approaching the Mott insulator despite the increasing
e-ph coupling.
The phonon distribution remains centered around $X=0$ and is barely
distinguishable from that of uncoupled phonons.
In other words the presence of $J$ completely quenches the phonon
degrees of freedom at least for these parameters.

We have shown that the competition between $\lambda$ and $J$ can lead
to two completely different strongly correlated phases for large
$U$. In the absence of $J$ we find a low-spin Hubbard-bipolaronic
state, while the prevalence of $J$ over $\lambda$ leads to a high-spin
Mott state without phononic signatures.
This confirms the expectation that a JT phonon mode only competes with
the Hund's coupling\cite{Capone_Science,Nomura2015,Nomura2016}, while
its effect can coexist with Mott localization. We can understand and
rationalize this result  by integrating out the JT mode. This leads to
a retarded (frequency-dependent) electron-electron interaction
$U^{e-ph}(\omega) = \frac{2\omega_0 g^2(n_1-n_2)^2}{\omega^2
  -\omega_0^2}$.
In the antiadiabatic limit $\omega_0 \gg D$ this becomes a static
effective interaction that adds up to the electron-electron interaction
\begin{align}
\label{eq:retarded}
U^{eff}=- & \lambda  (n_1-n_2)^2 = -\lambda\sum_{i,a} n_{ia\uparrow}n_{ia\downarrow} + \nonumber\\
+& \lambda \sum_{i,a<b, \, \sigma} n_{ia\sigma}n_{ib\sigma}
 + \lambda\sum_{i,a\neq b} n_{ia\uparrow}n_{ib\downarrow}
\end{align}
\begin{figure}
    \centering
    \includegraphics[width=1\linewidth]{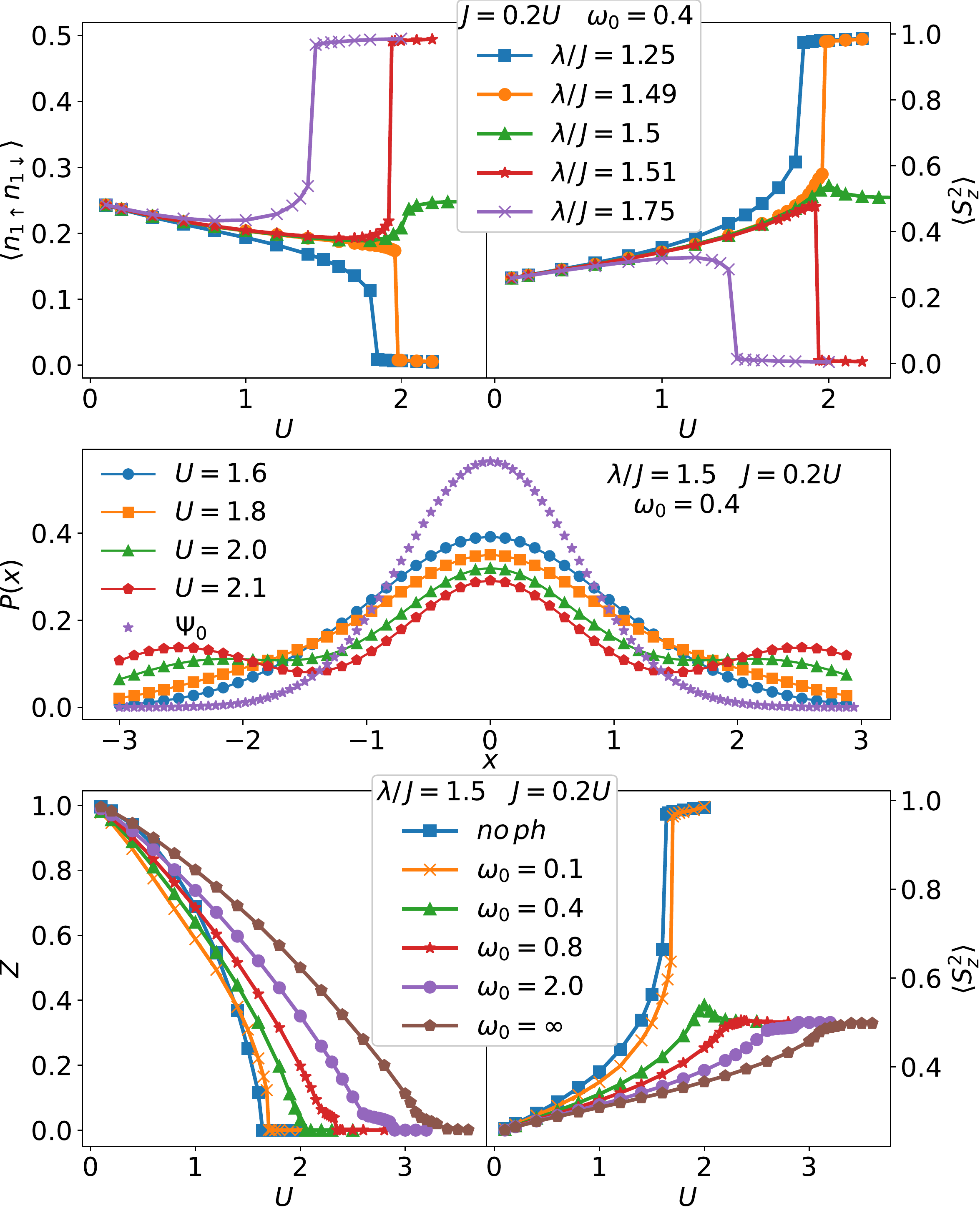}
    \caption{Top panels: Evolution of the intraorbital double occupations (left)
      and local spin (right) as a function of $U/D$ for different
      values of $\lambda/J$ and fixed $J/U =0.2$.
      Middle panels: phonon distribution function $P(X)$ for
      $\lambda/J = 3/2$ for different $U/D$.
      Bottom panels: Evolution of $Z$ and $\langle S^2_{z}\rangle$ as function
      of $U$ for different values of $\omega_0/D$ with $\lambda/J=3/2$.
    }
    \label{fig:3}
\end{figure}
The intra-orbital repulsion is thus reduced from $U$ to $U-\lambda$, while the inter-orbital interactions are enhanced to 
$U-2J+\lambda$ (opposite spins) and $U-3J+\lambda$ (parallel
spins). Interestingly, this opens the opportunity for an interorbital repulsion larger than the intraorbital one, a situation which is never found
in a purely electronic model.
We can expect that the high-spin Mott insulator (bipolaronic Mott) will be
realized when the effective intra-orbital (inter-orbital) term preveals on the other.
The boundary between the two phases should thus correspond to $U - \lambda
\simeq U - 3J + \lambda$ or $\lambda \simeq \frac{3}{2}J$.

In Fig.~\ref{fig:3} we show DMFT results for the region around
$\lambda/J = 3/2$ with $J/U = 0.2$ and $\omega_0/D=0.4$.
We see that the system evolves towards the high-spin
Mott insulator (zero double occupancy, high-spin) for $\lambda/J <
3/2$, while it converges towards the Mott bipolaron (large double
occupancy, low-spin) for $\lambda/J > 3/2$, in agreement with the
above estimate.
Precisely at $\lambda/J = 3/2$ we find
a superposition between the two above solutions where the
observables tend to the average of the limiting insulators.
For instance, the double occupancy converges towards 1/4 while the
local spin reaches 1/2.
We checked that this {\it hybrid} Mott state features both the local
configurations of the bipolaronic insulator ($\vert
  2,0\rangle$ and  $\vert 0,2\rangle$) and
that of the Mott insulator ($\vert 1,1\rangle$) with equal weights. 
The nature of the hybrid Mott insulator is further clarified
by the behavior of $P(X)$.
Increasing $U$ and $\lambda$ the phonons develop a
{\it {trimodal}} distribution with maxima at $X=0$ and at $X = \pm
X_0$,
i.e. a superposition of the distributions of the two
paradigmatic insulators.

We notice that, unlike similar systems described by a multiorbital
Hubbard model which show a Hund's metal solution at the boundary between two
insulators\cite{Isidori2019,Richaud}, the hybrid solution bridging
between two paradigmatic insulators is not metallic.
The reason is that all the local configurations characterizing the
hybrid state have the same occupation and they can not be connected by
hopping processes.

\begin{figure}
    \centering
    \includegraphics[width=1\linewidth]{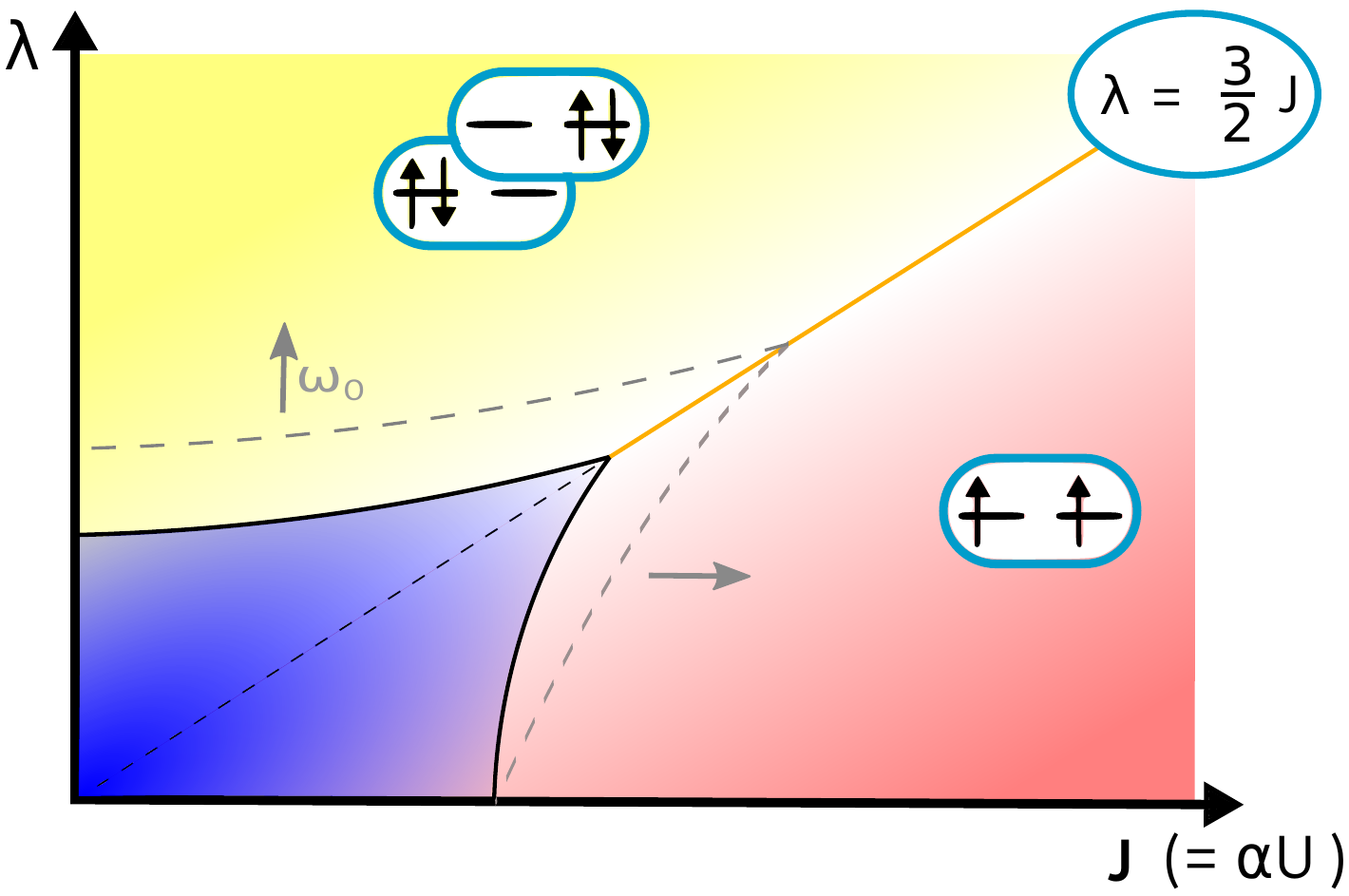}
    \caption{Schematic phase diagram of our model in the
      $U$-$\lambda$-$\omega_0$ space.
      The yellow region is the Mott-bipolaronic state, while the red
      region is the high-spin Mott insulator.
      The blue area is the metallic solution.
      Increasing $\omega_0$ the metallic region extends as shown by the arrows and the grey dashed lines.
    }
    \label{fig:4}
\end{figure}
We finally discuss the role of the phonon frequency $\omega_0$ (bottom
row of Fig.~\ref{fig:3}).
The hybrid insulator is indeed realized when the phonon frequency
exceeds a threshold of the order $\omega_0/D =0.2$.
For smaller $\omega_0$ the system becomes similar to the purely
electronic model. It is however interesting that the simple analytical
estimate obtained in the antiadiabatic limit agrees with numerical
data also down to $\omega_0/D \ll 1$, suggesting that retardation
effects come into play only close to the limit $\omega_0 \to
0$. Quantitatively, we found that the critical $U$ for the hybrid Mott
state is decreased by reducing $\omega_0/D$. 

Our main conclusions can be summarized in the schematic phase diagram of
Fig.~\ref{fig:4} in $J$-$\lambda$ plane (with $U$ proportional to $J$) where a metallic phase is stable when both interactions are small. 
The strong-coupling solution largely depends
on the ratio $\lambda/J$ betweeen the e-ph and the Hund's couplings.
For small $J$ we find a bipolaronic-Mott insulator (yellow region)
where the Mott localization and the tendency to form bipolarons
coexist and cooperate.
For small $\lambda$ we have a high-spin Mott insulator
(red region) where the phonon fingerprints are
washed out by the Hund's coupling.
Setting $\lambda/J = 3/2$ we reach a hybrid Mott insulator which shows
features of both the above insulators as shown by a peculiar trimodal
phonon distribution. Increasing the phonon frequency enlarges the
metallic region. 

Our results highlight that the interplay between strong electronic correlations and phonons can be subtle and multifaceted, in contrast with the HH model where the two effects are essentially exclusive. In particular we can have regimes where the Mott transition is favoured by phonons and phononic fingerprints are clear in the strongly correlated metal and in the Mott insulators, as well as regime where the effect of phonons is completely quenched. The intermediate regime where we have a superposition between the two insulators is particularly intriguing even if it is found only for a specific line of the phase diagram. This information can be used as a building block to understand the properties of strongly correlated materials with multiorbital electronic structure and coupling to phonons beyond the
limiting paradigm of the single-band Hubbard-Holstein model. 

\label{sec:Acknowledgements}
We acknowledge financial support from MIUR through the PRIN 2017
(Prot. 20172H2SC4 005) programs and Horizon 2020 through the ERC
project FIRSTORM (Grant Agreement 692670).

\bibliographystyle{apsrev4-1}
\bibliography{biblio}

\end{document}


\title{Supplementary Material - Competing Correlated Insulators in multi-orbital systems coupled to phonons}

\author{Alberto Scazzola}
\affiliation{Scuola Internazionale Superiore di Studi Avanzati (SISSA), Via Bonomea 265, I-34136, Trieste, Italy}
\author{Adriano Amaricci}
\affiliation{CNR-IOM Democritos, Via Bonomea 265, I-34136 Trieste, Italy}
\author{Massimo Capone}
\affiliation{Scuola Internazionale Superiore di Studi Avanzati (SISSA), Via Bonomea 265, I-34136, Trieste, Italy}
\affiliation{CNR-IOM Democritos, Via Bonomea 265, I-34136 Trieste, Italy}
\date{\today}

\maketitle

\section{Model and method}

We consider a two-band Hubbard model on an infinite coordination Bethe lattice, with a semicircular density of states for each band

\begin{equation}
    \mathcal{D}(\epsilon)=\frac{2}{\pi D^2}\sqrt{D^2-\epsilon^2}, \quad |\epsilon|<D
\end{equation}

where $D$ is the half-bandwidth that we used as energy scale in this work. The two bands are degenerate in order to focus only on the local interaction terms of the model Hamiltonian. 

There are two groups of interaction terms: electron-electron interaction $H_{e\--e}$ and electron-phonon interaction $H_{e\--ph}$. The first one features the local repulsion terms typical of the Kanamori Hamiltonian, except the spin-flip and pair-hopping terms that we omitted, since they don't change qualitatively the main picture described in this work.

\begin{align}
\label{eq:H_ee}
H_{\mathrm{e-e}} & = U\sum_{i,a} n_{ia\uparrow}n_{ia\downarrow} 
+ (U'-J)\sum_{i,a<b, \, \sigma} n_{ia\sigma}n_{ib\sigma} \nonumber\\
& + U'\sum_{i,a\neq b} n_{ia\uparrow}n_{ib\downarrow}  
\end{align}

where $i$, $a$ and $b$, $\sigma$ are respectively site, orbital and spin indices. Here we use the standard condition $U'=U-2J$, that connects the inter- ($U'$) and intra-orbital ($U$) repulsions and Hund's coupling ($J$) and enforces orbital rotation invariance\cite{PhysRevB.18.4945} 

The second group of interaction terms contains a single dispersionless bosonic mode with frequency $\omega_0$ locally coupled to the occupation difference between the electronic orbitals via the coupling constant $g$. This is one of the two Jahn-Teller modes active in a two-orbital $e_g$ manifold, usually denoted as $Q_3$ (the other one, $Q_2$, being coupled to the local hybridization between the orbitals).

\begin{align}
H_{\mathrm{e-ph}} & = g\sum_{i\sigma} (n_{i1\sigma} -n_{i2\sigma})(a_i^{\dagger} + a_i) + \omega_0 \sum_{i} a_i^{\dagger}a_i
\end{align}

In the following we will always deal with the electron-phonon interaction referring to the parameter $\lambda=2g^2/\omega_0$ that describes the strength of the effective electron-electron interaction mediated by the phonons, as can be showed integrating out the phononic degrees of freedom from the action of our model.

The density of the system has been set by the half-filling condition, i.e. two electrons per site $n=2$, in order to observe the Mott localization and its interplay with the phonon mode.

We solve the model using Dynamical Mean Field Theory (DMFT)\cite{DMFT} and the related impurity problem using an exact diagonalization solver\cite{amaricci2021edipack} at zero temperature with $n_b=3$ bath sites for each orbital, leading a total number of levels $N_b=2(1+n_b)=8$. A cutoff on the maximum number of phonons is also needed in order to have a finite Hilbert space and we set it to $N^{ph}_{max}\sim 25$. This cutoff is adjusted depending on the parameters of the Hamiltonian, in particular reducing $\omega_0$ requires to increase $N^{ph}_{max}$ since the excitation of phonons is easier.

\section{Observables and details}

There are several observables we can calculate to collect information about the interplay between the different interaction terms. Here we report the observables shown in the main text of the paper together with some others that add more details to our picture. 

The quasiparticle weight $Z=(1-\partial\mathfrak{I}\Sigma/\partial\omega_n|_{\omega_n\rightarrow 0})^{-1}$, with $\Sigma$ the self-energy and $\omega_n$ the Matsubara frequencies, is related to the inverse of the effective mass of the electrons in a Fermi liquid picture. A vanishing $Z$ characterizes a strongly correlated insulator, while $Z=1$ is obtained in the non-interacting limit.

The intra-orbital double occupation $\langle n_{im\uparrow}n_{im\downarrow}\rangle$ and the local magnetic moment measured by $\langle S^2_z\rangle = \langle (\sum_m S_{zim})^2\rangle$ are key observables to discern between the two different insulating phases we found with this model. The Mott-bipolaronic insulator is characterized by a vanishing local magnetic moment and a high intra-orbital double occupation, indicating the presence of an effective intra-orbital electron-electron attraction, while the high-spin Mott one shows the opposite behavior. Here we also present the inter-orbital charge correlation $\langle n_1n_2\rangle-\langle n_1\rangle\langle n_2\rangle$. It is well known that the Hund's coupling favours a homogeneous occupation of the orbitals and this causes the freezing of orbital degrees of freedom, leading to an effective decoupling in some regimes. Indeed, we find a vanishing charge inter-orbital correlation in the high-spin Mott insulator. In the case of the Mott-bipolaronic one, instead, this observable is enhanced by the phonons and the orbitals remain strongly correlated even in the insulating phase.

\begin{figure}[t]
    \centering
    \includegraphics[width=1\linewidth]{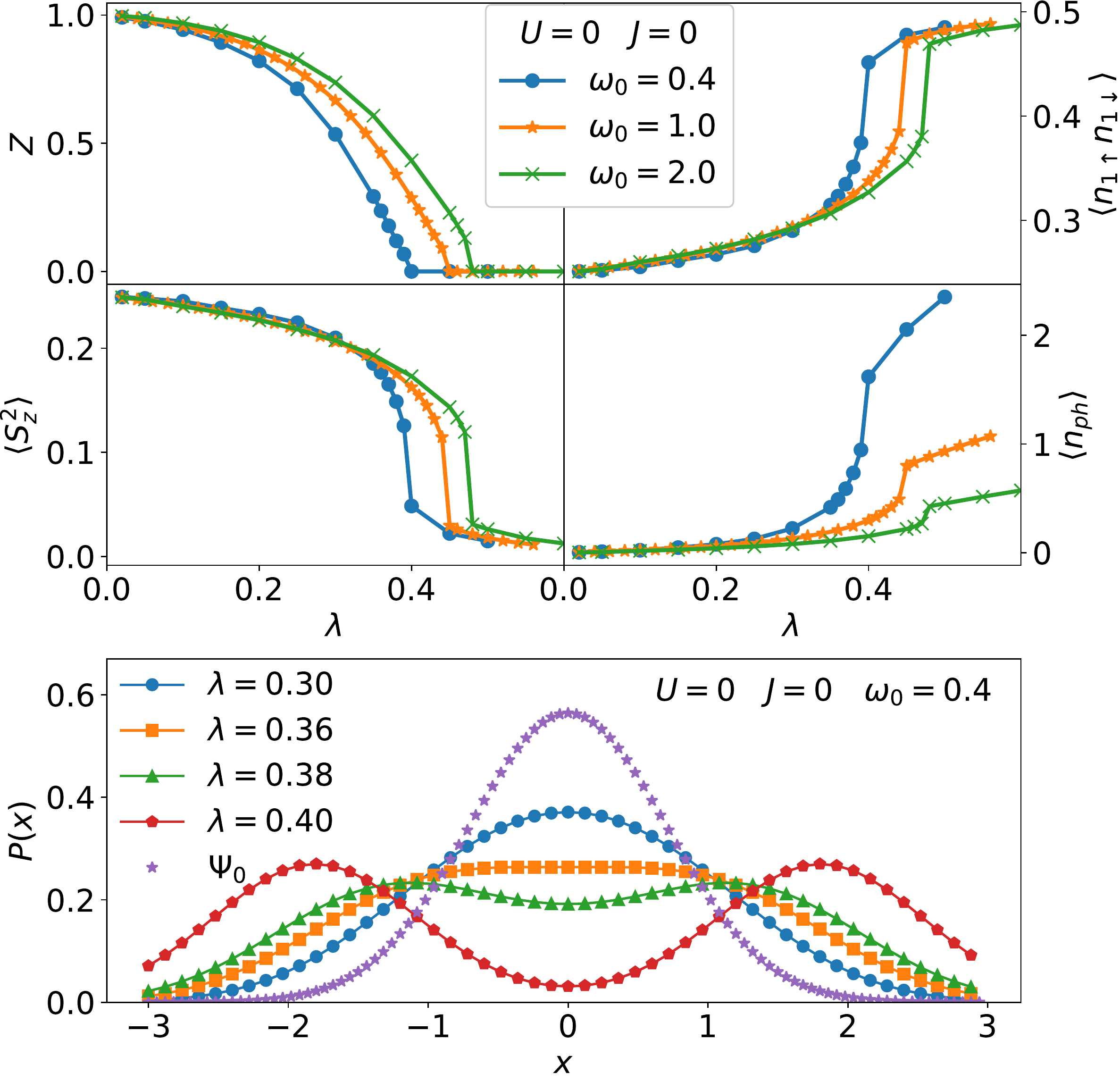}
    \caption{Summary of the results with $U=J=0$. Top: main observables as a function of $\lambda$ for three different $\omega_0$. Bottom: phonon distribution function across the metal-insulator transition with $\omega_0=0.4$.}
    \label{fig:0}
\end{figure}

We also consider three observables related to the phonon degrees of freedom: the average number of excited phonons $N_{ph}=\langle a^\dag a\rangle$, the renormalized phonon frequency $\Omega$ and the phonon distribution function $P(X)$. $\Omega$ is computed from the phonon Green's function $D_{ph}$ as 

\begin{equation*}
    \Omega = \sqrt{-\frac{2\omega_0}{D_{ph}(i\omega_n=0)}}
\end{equation*} 

and we use it to study the softening of the phonon mode \cite{Capone2003p}. $P(X)$ is the the probability for the phonon displacement operator to assume a specific value. It is computed as

\begin{equation*}
    P(X) = |\langle \Phi_0|X\rangle|^2 = \sum_{nm}\psi_n(X)\psi_m(X)\langle \Phi_0|n\rangle\langle m|\Phi_0\rangle
\end{equation*}

where $|\Phi_0\rangle$ is the ground state of the system and $\psi_n(X)$ are the eigenfunctions of the harmonic oscillator. When the phonons are decoupled from the electrons this is nothing more than a gaussian centered around $X=0$, while it becomes a bimodal distribution close to the transition between metal and bipolaronic insulator, signaling the so called polaron crossover. This modification of the distribution is related to a spatial distortion of the lattice \cite{Capone2006p}.

\subsection{Pure electron-phonon interaction}

The first case we consider involves the electron-phonon interaction only, i.e. $U=J=0$. As described in the main text of the paper, $Z$ decreases as a function of $\lambda$, leading to a metal-insulator transition. The insulating phase is the so called bipolaronic insulator, with high intraorbital double occupancy, vanishing local magnetic moment and a meaningful number of excited phonons. Close to the transition the phonon distribution function becomes bimodal (bottom panel of Fig.\ref{fig:0}), signaling the polaron crossover and the subsequent onset of the bipolaronic insulating state.

The picture described is not influenced qualitatively by the value of the phonon frequency as showed in the top panels of Fig.\ref{fig:0}: increasing $\omega_0$, the metal-insulator transition is pushed towards higher $\lambda$ and the number of excited phonons gets reduced as $\sim 1/\omega_0$, but the final insulating phase is always the same.

\begin{figure}[b]
    \centering
    \includegraphics[width=1\linewidth]{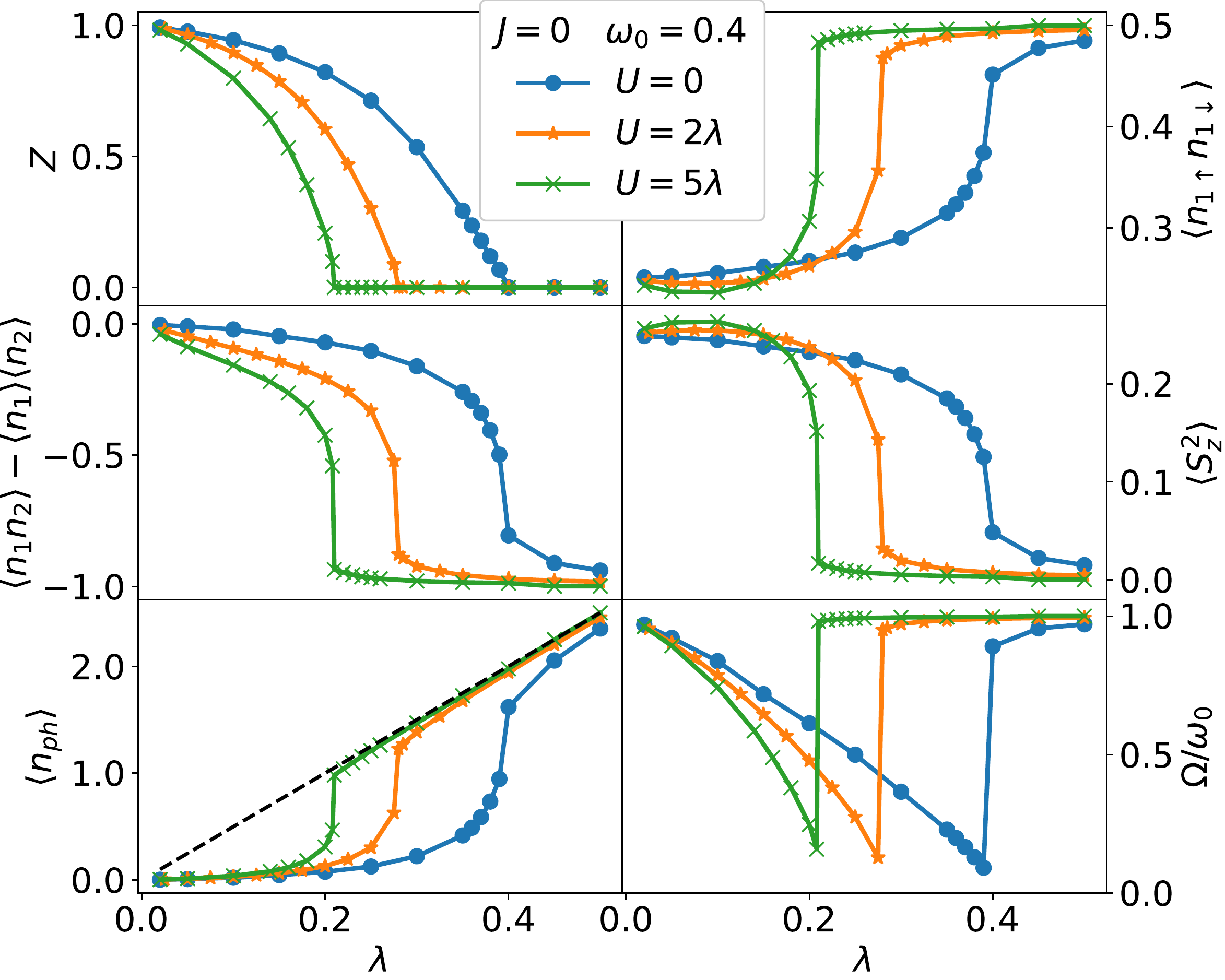}
    \caption{Evolution of the observables as a function of $\lambda$ with $J = 0$ and fixed $U/\lambda$ for $\omega_0/D = 0.4$. The black dotted line in the panel of $n_{ph}$ represents the analytic result in the atomic limit.}
    \label{fig:1}
\end{figure}

\subsection{Zero Hund's coupling}

We now add the Hubbard repulsion, but we still neglect the Hund's coupling. The observables of Fig.\ref{fig:1} show that the insulating phase we find is always the same regardless of the value of the ratio $U/\lambda$, i.e there is no competition between $\lambda$ and $U$. The insulator is a Mott-bipolaronic one, with high intra-orbital double occupation and vanishing local magnetic moment. The inter-orbital charge correlation shows that the two orbitals are strongly correlated inside the insulating phase. The local electronic state of the system is indeed a superposition of orbitally polarized configurations, with an empty orbital and a full one. Therefore the occupation of one orbital is related to the occupation of the other.

\begin{figure}[b]
    \centering
    \includegraphics[trim={0.5cm 0.5cm 0cm 0.5cm},clip, width=0.7\linewidth] {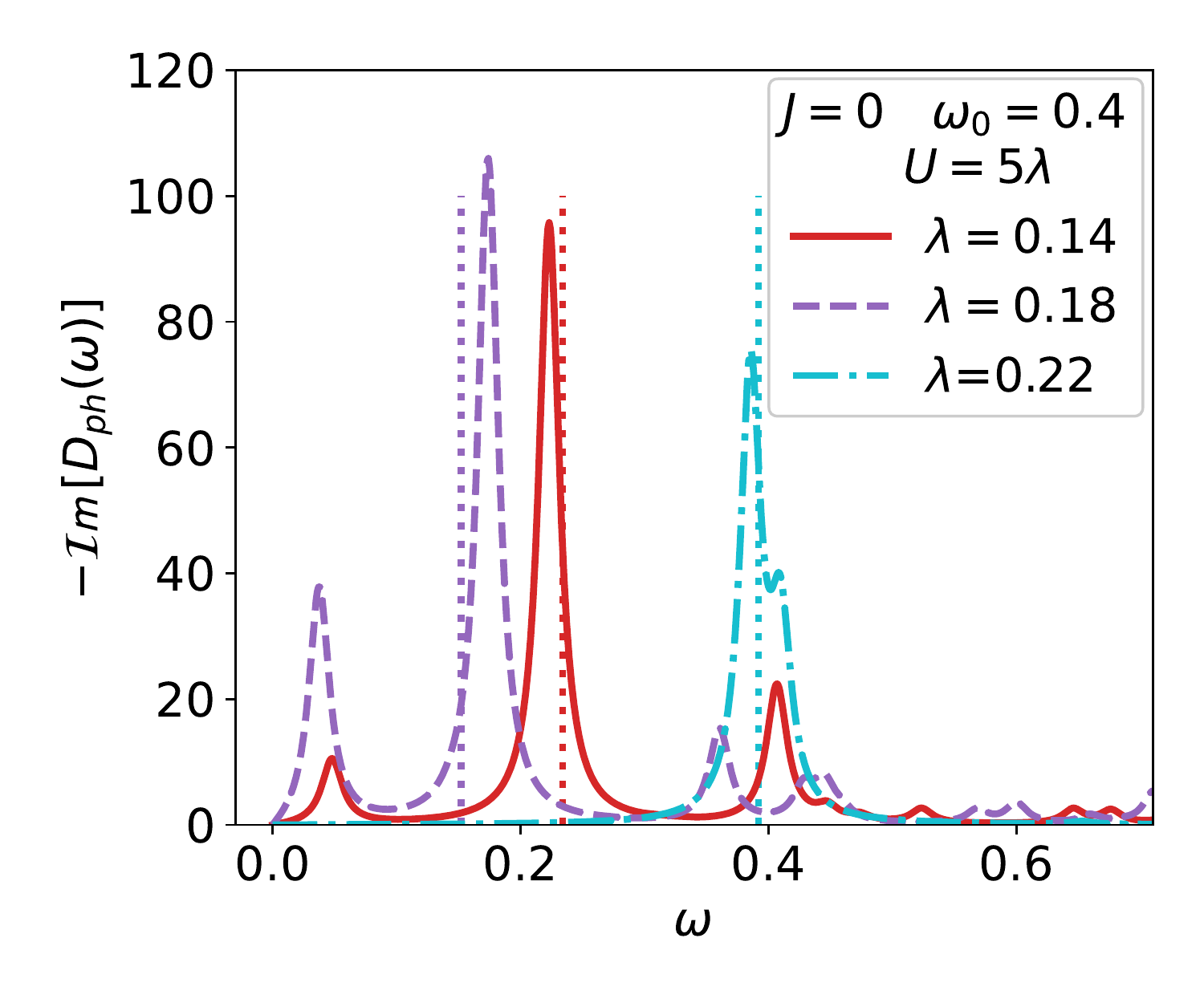}
    \caption{Imaginary part of the real-frequency phonon Green's function across the metal-insulator transition. The vertical dotted lines represent the corresponding renormalized frequencies.}
    \label{fig:2}
\end{figure}

The number of phonons increases rapidly close to the transition and then it becomes linear in $\lambda$. The reason behind this linear behavior is the freezing of the charge degrees of freedom inside the insulating phase, so that the electronic operators don't contribute anymore to the variation of the number of phonons. This becomes clear considering the Hamiltonian of the system without the kinetic term (atomic limit): the model is exactly solvable and the number of phonons can be calculated as

\begin{equation*}
\begin{aligned}
    N_{ph}=&\frac{\lambda}{2\omega_0}\langle(n_1-n_2)^2\rangle=\\
          =&\frac{\lambda}{2\omega_0}(\langle n\rangle+2\langle n_{1\uparrow}n_{1\downarrow}\rangle+2\langle n_{2\uparrow}n_{2\downarrow}\rangle-2\langle n_{1}n_{2}\rangle)
\end{aligned}
\end{equation*}

All the expectation values inside this formula are constant in the insulator, leading to a linear dependence of $N_{ph}$ on $\lambda$, showed by the black dotted line in the corresponding panel of Fig.\ref{fig:1}. We expect this result to be meaningful inside the insulating phase when the kinetic part of the Hamiltonian is negligible.

The last panel of Fig.\ref{fig:1} shows the renormalized phonon frequency, with a strong softening of the mode when getting closer to the transition and then a sudden jump. Fig.\ref{fig:2} is useful to understand the meaning of $\Omega$ and its behavior: here we show the imaginary part of the real-frequency phonon Green's function $D_{ph}(\omega)$ and the corresponding renormalized frequencies (vertical dotted lines). When the interaction parameters are increased the spectrum is shifted towards lower energy and this behavior is captured by the reduction of the renormalized frequency, which represents some sort of weighted average of the peaks in the Green's function. After the metal-insulator transition, the spectral weight suddenly moves to an energy $\sim \omega_0$, resembling the result expected for a decoupled mode. The reason for the jump is the same as that for the linear behavior of $N_{ph}$, i.e. the freezing of the charge degrees of freedom inside the insulating phase.

\begin{figure}[t]
    \centering
    \includegraphics[width=1\linewidth]{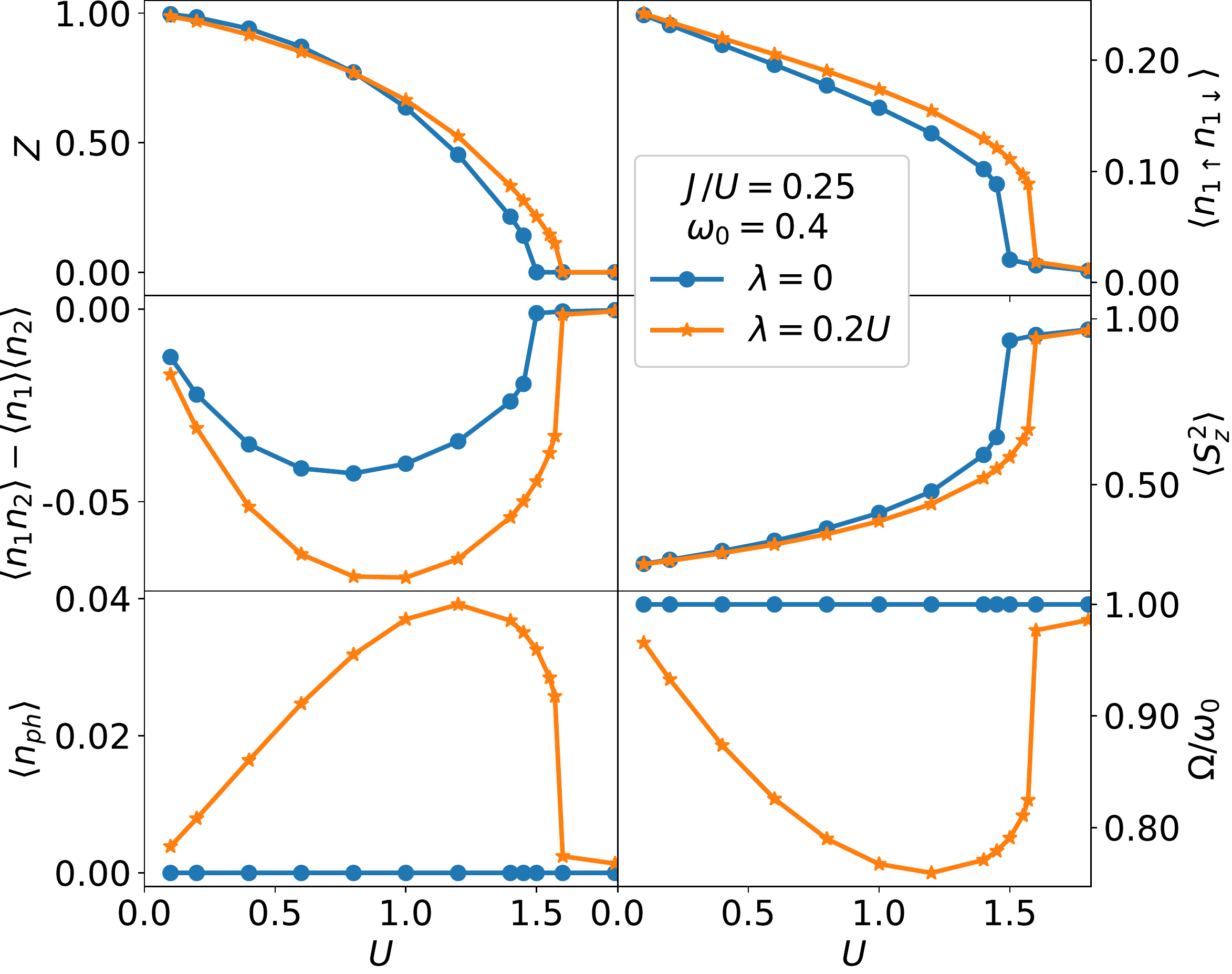}
    \caption{Evolution of the observables as a function of $U$ with $J/U = 0.25$ and either $\lambda=0$ or $\lambda/U=0.2$ for $\omega_0/D = 0.4$.}
    \label{fig:3}
\end{figure}

\subsection{Small electron-phonon coupling}

Here we consider a completely different scenario, adding a finite Hund's coupling $J$ to the model with either no phonons or a small electron-phonon coupling. We thus find a different insulating phase characterized by local electronic configurations featuring one electron per orbital with parallel spin. This causes the presence of high magnetic moments and vanishing intra-orbital double occupation. We will refer to this phase as a high-spin Mott insulator. The summary of several observables as a function of $U$ is presented in Fig.\ref{fig:3}, where we keep $J/U$ and $\lambda/U$ fixed. The addition of a small $\lambda$ shifts the transition towards higher $U$, signaling the presence of a competition with $J$.

The inter-orbital charge correlation shows a non-monotonic behavior: when the interaction parameters are small, the degree of correlation increases because the main effect on the system is caused by $U$, since $J$ is too small. Then the presence of the Hund's coupling becomes important causing a drop of the inter-orbital charge correlation, that vanishes at the transition: the charge degrees of freedom of the two orbitals are effectively decoupled in the insulating phase. This is the contrary with respect to the Mott-bipolaronic insulator.

The number of phonons and the renormalized frequency show a different behavior with respect to Fig.\ref{fig:1}. The first one increases up to a maximum, that is anyway very small, and then rapidly disappear close to the transition. The second one shows only a moderate softening and no sudden jump close to the transition. This results confirm a suppression of the phonons when the Hund's coupling is stronger than the electron-phonon one.

\begin{figure}[t]
    \centering
    \includegraphics[width=1\linewidth]{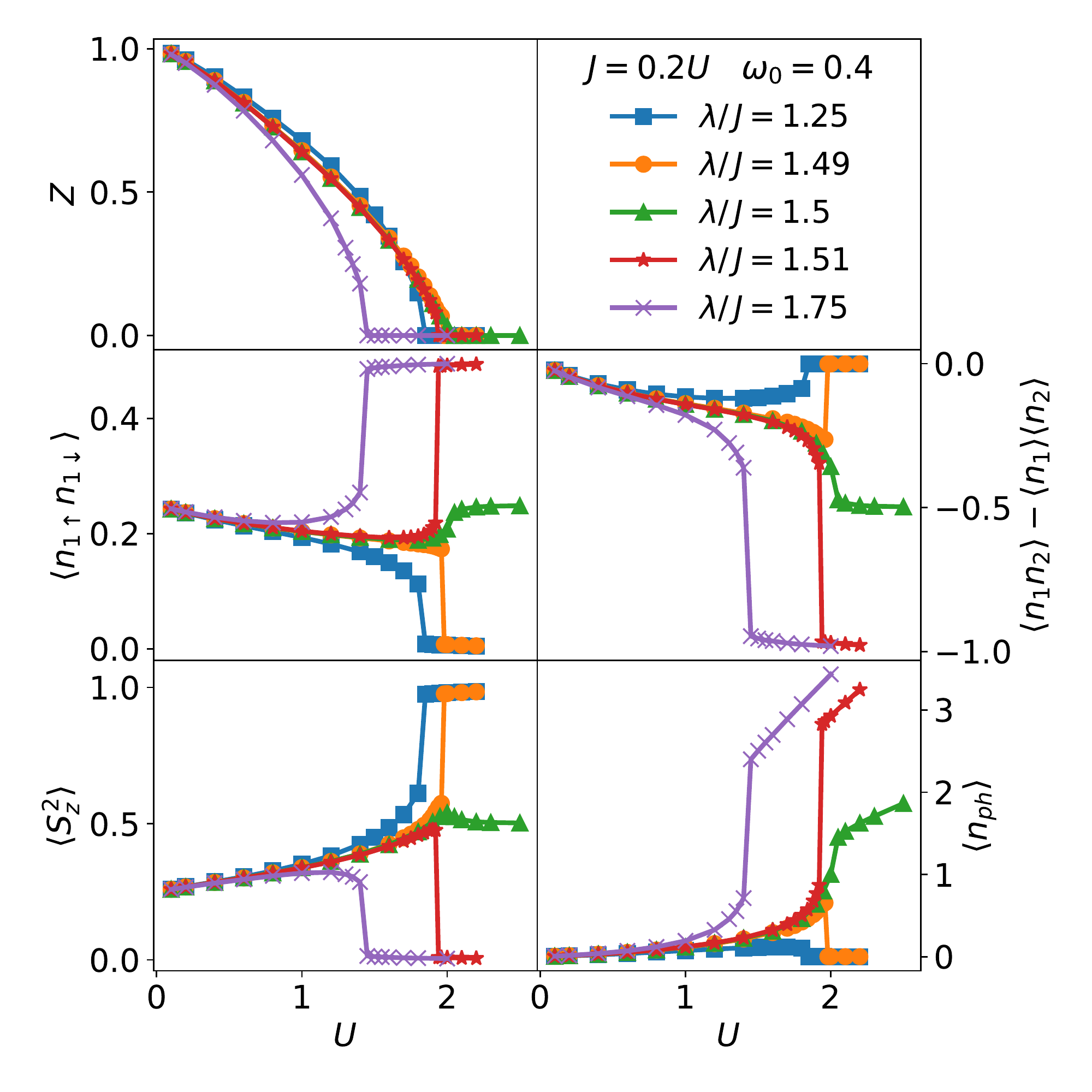}
    \caption{Evolution of the observables as a function of $U$ with $J/U = 0.2$ and $\omega_0/D = 0.4$ for different values of the fixed ratio $\lambda/J$.}
    \label{fig:4}
\end{figure}

\subsection{Competition between $\lambda$ and $J$}

\begin{figure}[b]
    \centering
    \includegraphics[trim={0.5cm 0.6cm 0.5cm 0.6cm},clip,width=0.8\linewidth] {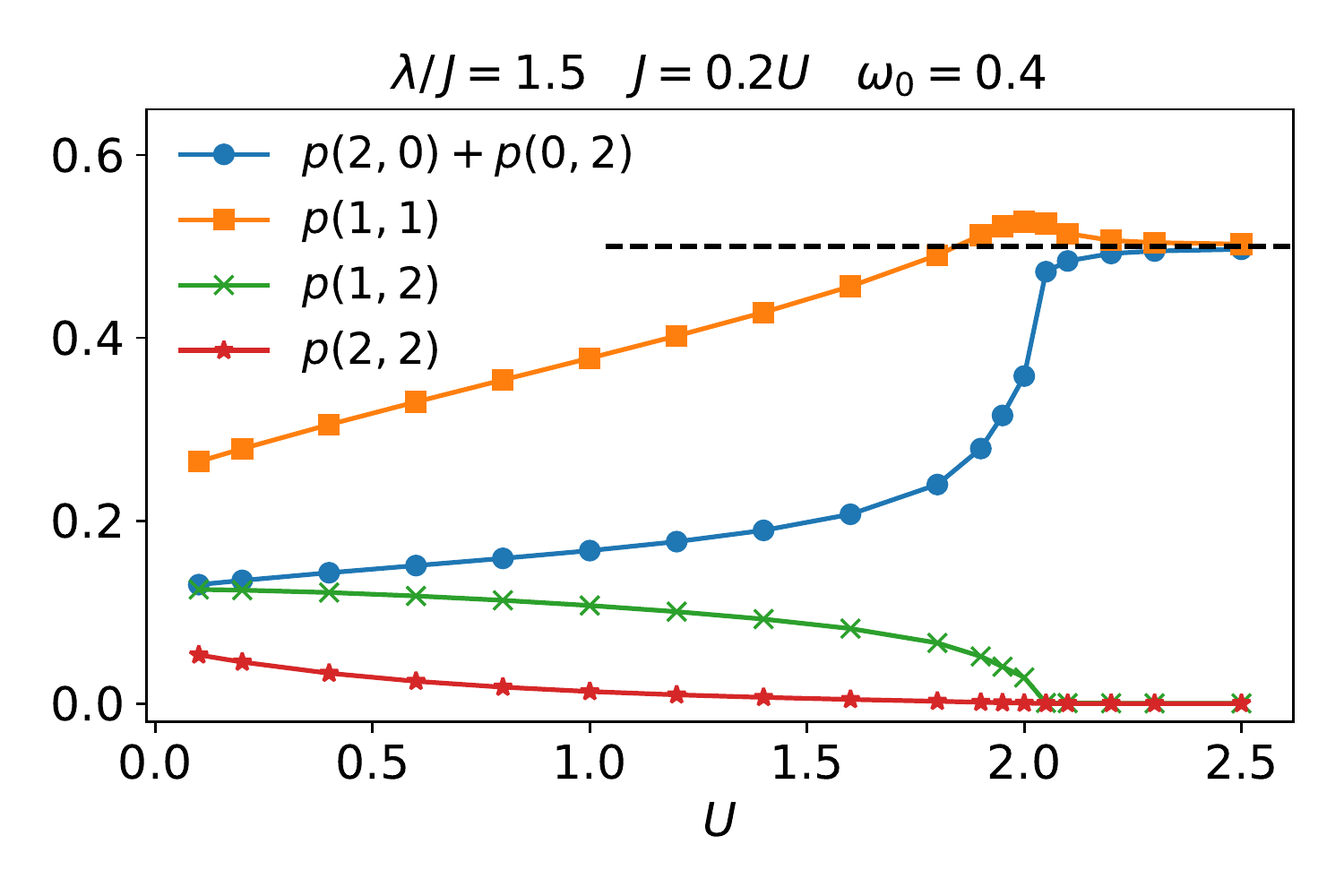}
    \caption{Occupation probability of different local electronic configurations with $J/U = 0.2$, $\lambda/J=1.5$ and $\omega_0=0.4$. The horizontal dotted line corresponds to $0.5$.}
    \label{fig:5}
\end{figure}

In this last section we explicitly collect results about the competition between $\lambda$ and $J$. The evolution of several observables is showed in Fig.\ref{fig:4} considering various values of the ratio $\lambda/J$ around the condition of "strongest competition", as discussed in the main text of the paper. The two different insulating phases are separated by a very sharp line at $\lambda/J=1.5$, where the system is found in a mixed phase: the two insulators are degenerate and coexist. This mixed insulating phase features a situation in which half of the lattice sites will show a local electronic configuration compatible with the bipolaronic-Mott insulator while the other half with the high-spin one. For this reason every observable assumes a value that is the average of those found for the two insulators. The degeneracy characterizing this insulating phase becomes clear looking at the occupation probability of each local electronic configuration (see Fig.\ref{fig:5}). We refer to these probabilities as $p(n,m)$, where $n$ and $m$ describe respectively the occupation of orbital 1 and orbital 2, without any distinction of the spin configuration. For this reason $p(1,1)$, for example, will contain the contribution from 4 different local spin configurations, while $p(2,0)$ only one. Given this definition, there are 4 independent occupation probabilities due to the half-filling condition and the degeneracy of the orbitals. In the insulating phase $p(1,1)$ and $p(0,2)+p(0,2)$ are both equal to 0.5 as expected. Notice that we have to compare the sum of the orbital polarized states with $p(1,1)$ because this last probability contains the contribution from two spin configuration, $(\uparrow,\uparrow)$ and $(\downarrow,\downarrow)$.

\begin{figure}[t]
    \centering
    \includegraphics[width=1\linewidth]{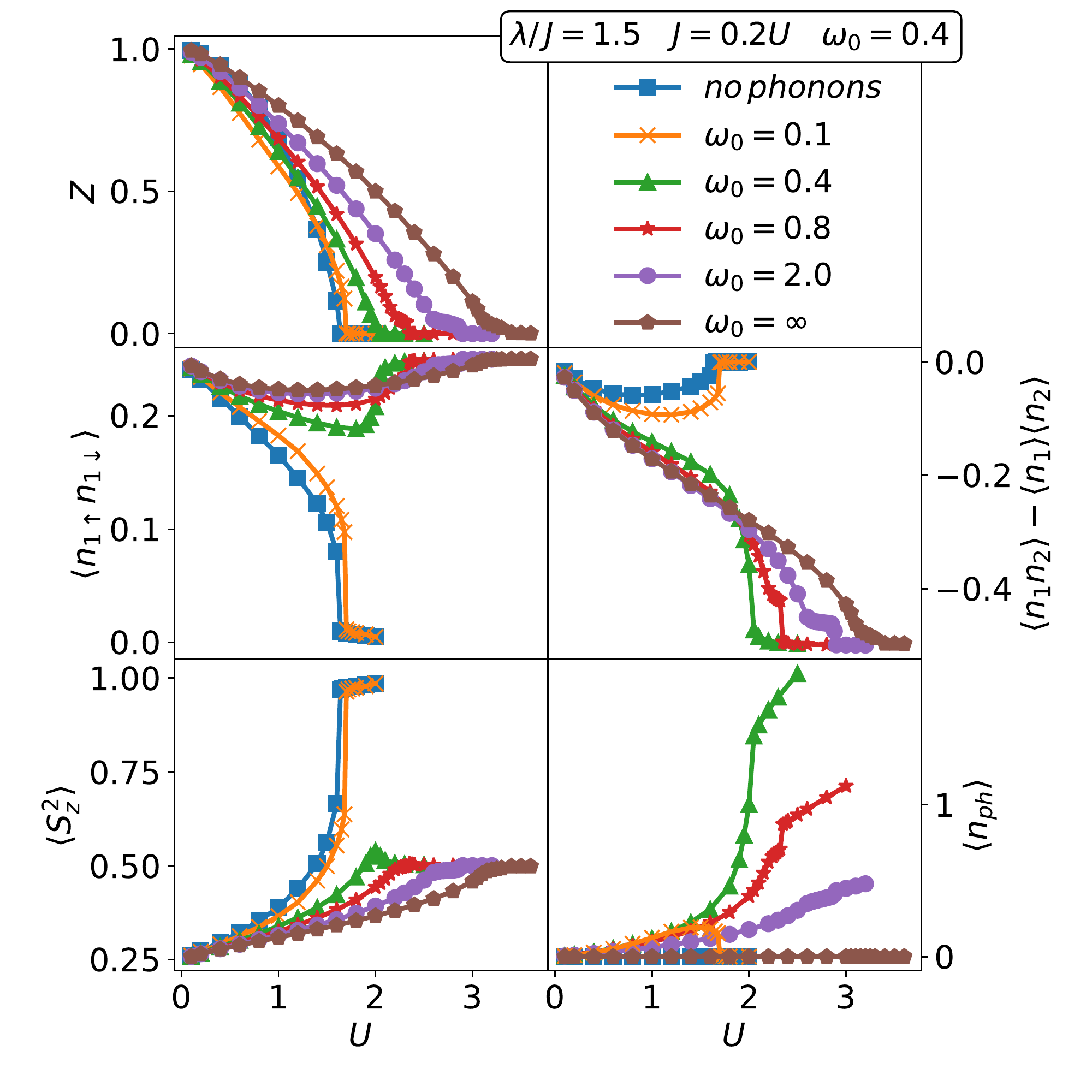}
    \caption{Evolution of the observables as a function of $U$ with $J/U = 0.2$ and $\lambda/J=1.5$ for several values of $\omega_0$.}
    \label{fig:6}
\end{figure}

The mixed phase is interesting because it allows to study the effect of other parameters on the competition between $\lambda$ and $J$, since the fragile balance allowing the existence of this insulating phase can be easily broken by perturbations acting either on the phononic or the electronic part.
For this reason we analyse the stability of the results with respect to the variation of the phonon frequency focusing on the $\lambda/J=1.5$ line.

The observables presented in Fig.\ref{fig:5} show that reducing $\omega_0$ weakens the effect of the phonons up to the point where there is a break down of the mixed insulating phase and the system transitions to the high-spin Mott insulator. There are two clear limiting cases between which all the electronic observables evolve for each finite $\omega_0$: the first one is the anti-adiabatic limit, i.e. $\omega_0\rightarrow\infty$, and the second one is the absence of phonons. In the first one the effective retarded interaction mediated by the phonons becomes instantaneous and the system is described by a purely electronic model with renormalized interaction parameters (see main text of the paper). The second limiting case is realized for very small phonon frequency, when the retardation effects become dominant, and we find that the system behaves as if there were no phonons at all.

\bibliographystyle{apsrev4-1}
\bibliography{biblio}